\documentclass[prd,nofootinbib,showpacs,superscriptaddress,preprint]{revtex4}
\usepackage[T1]{fontenc}
\usepackage{amsmath,amssymb}
\usepackage{epsfig}
\usepackage{graphicx}
\usepackage[usenames,dvipsnames]{color}
\usepackage{slashed}
\usepackage[colorlinks,citecolor=blue]{hyperref}
\usepackage{pdfpages}
\usepackage{color}
\begin{document}
\title{Normal hierarchy neutrino mass model revisited with leptogenesis}
\author{Ananya Mukherjee}
\email{ananyam@tezu.ernet.in}
\author{Mrinal Kumar Das}
\email{mkdas@tezu.ernet.in}
\author{Jayanta Kumar Sarma}
\email{jks@tezu.ernet.in}
\affiliation{%
 Department of Physics, Tezpur University, Tezpur 784\,028, India %
 }%

\begin{abstract}
We have studied the scenario of baryogenesis via leptogenesis in an $A_4$ flavor symmetric framework considering type I seesaw as the origin of neutrino mass. Because of the presence of the fifth generation right handed neutrino the model naturally generates non-zero reactor mixing angle. We have considered two vev alignments for the extra flavon $\eta$ and studied the consequences in detail. As a whole the additional flavon along with the extra right handed neutrinos allow us to study thermal leptogenesis by the decay of the lightest right handed neutrino present in the model. We have computed the matter-antimatter asymmetry for both flavor dependent and flavor independent leptogenesis by considering a considerably wider range of right handed neutrino mass. Finally, we correlate the baryon asymmetry of the universe (BAU) with the model parameters and light neutrino masses. 
\end{abstract}

\maketitle

\section{Introduction}
\label{sec1}
The Standard Model(SM)with its enormous success stands at the top after the milestone discovery of a 125 GeV neutral scalar boson, \textit{the} Higgs Boson. Still, there remained some unanswered phenomena in this ballpark of particles and forces like matter-antimatter asymmetry, origin of neutrino mass and the observed dark matter, which keeps the window open for physics beyond Standard Model. It will be compelling if we can bring all of these occurrences into a single frame. With this motivation we put forward a theory which addresses the neutrino phenomenology consistent with recent global fit oscillation data along with a possible origin of matter-antimatter asymmetry. 

 \begin{table}[htb]
        \centering
       \begin{tabular}{|c|c|c|}
       \hline  Oscillation parameters   & $3\sigma$(NO)&  $3\sigma$(IO) \\ 
       \hline $\Delta \text{m}^{2}_{21} [10^{-5} eV^{2}] $  &  7.05 - 8.14  & 7.05 - 8.14 \\ 
       \hline $\Delta \text{m}^{2}_{3l}[10^{-3} eV^{2}] $  & 2.43 - 2.67 &  2.37-2.61\\ 
       \hline $ \text{Sin} ^{2}\theta_{12} $ & 0.273 - 0.379  & 0.273 - 0.379  \\ 
       \hline  $ \text{Sin} ^{2}\theta_{23}$ & 0.384 - 0.635 &   0.388 - 0.638   \\
       \hline  $ \text{Sin} ^{2}\theta_{13} $ & 0.0189 - 0.0239  & 0.0189 - 0.0239  \\
       \hline $\delta / \pi$ & 0 - 2.00 & 0.00-0.17 \& 0.79 - 2.00\\
       \hline
       \end{tabular}
     \caption{Latest global fit Neutrino Oscillation data \cite{deSalas:2017kay}} \label{tab1}       
           \end{table} 

The fact of existence of neutrino mass has been steadfastly established by the dedicated neutrino oscillation experiments \cite{Fukuda:1998mi,Ahmad:2002jz,arthur2016,kajita2016}. The status of neutrino oscillation data is presented in the table \ref{tab1}. The continuous rumble of various see-saw mechanisms \cite{seesaw} in the matter of explaining neutrino mass and their smallness in comparison to other fermions in the Standard Model is magnificent. Because of the absence of right handed neutrinos in the SM, neutrino mass is not explainable within this paradigm. Thus one has to go beyond Standard Model(BSM) by extending the fermion sector of the SM including two or more right handed neutrinos in order to implement see saw mechanism to make the theory viable for explaining neutrino mass and mixing. With robust number of evidences it has now become a proven fact that there exists tiny excess of matter over antimatter which is known as the Baryon asymmetry of the Universe(BAU). The dynamical process of production of baryon asymmetry from baryon symmetric era is familiar as baryogenesis. Although there are huge evidences that suggest the tiny excess of matter over antimatter that was produced in the early universe still its origin remains illusive. With this growing evidence there are several ways through which baryon asymmetry is realized. Among them baryogenesis via leptogenesis is considered as one of the most field theoretically consistent ways of explaining baryogenesis as proposed by Fukugita and Yanagida\cite{Fukugita:1986}. While realizing this picture it has come to notice that the L violating out of equilibrium decays of singlet neutrinos with mass larger than the critical temperature create an initial excess of lepton number L(for detail one may refer to \cite{leptogenesis}). This excess in lepton number is then partially converted into the baryon asymmetry of the universe(BAU) via (B+L) violating sphaleron transition \cite{sphaleron,Pilaftsis:1998pd,Flanz:1998kr}. 
 
 In the same context one can traditionally search for the possibility of foreseeing the lepton asymmetry generated in the lepton sector due to the presence of heavy right handed neutrinos(RHN) as the key ingredients for type I see-saw to take place. In this regard lot of works \cite{Borah:2017qdu,Borah:2014bda,Karmakar:2014dva} have been exercised for a common search, addressing these two problems within a single frame. In order to bring this scenario into picture, SM-singlet heavy RH neutrinos are introduced, which through a dimension five operator eventually gives rise to tiny Majorana neutrino masses. The lepton asymmetry is dynamically generated by the L violating out of equilibrium decay of the lightest RHN satisfying Sakharov's conditions (for detail please see \cite{Sakharov,Davidson:2008bu}) required for a nonzero baryon asymmetry of the universe. 

If we look at the Lagrangian for the leptons we see that it permits the lepton-number-violating decays of $\text{N}_i$ (for i = 1, 2) via: $\text{N}_i \rightarrow l + H^c$ and $\text{N}_i \rightarrow l^c + H$. Since each decay mode can take place at both tree and one-loop levels, the interference of two decay amplitudes contributes to a CP-violating
asymmetry $\epsilon _i$ between $\text{N}_i \rightarrow l +H^c$ and its CP-conjugated process $\text{N}_i \rightarrow l^c +H$. If Sakharov's third condition is satisfied the out-of-equilibrium decays of the lightest RHN $\text{N}_i$, $\epsilon _i$ may result in a net lepton number asymmetry which later on may convert into the observed predominance of matter over antimatter.
Such an elegant
baryogenesis-via-leptogenesis mechanism provides a viable interpretation of the cosmological baryon number asymmetry, which is a ratio of the difference in number densities of baryons ($\text{n}_B$) and anti baryons ($\text{n}_{\bar B}$)
to the entropy density of the universe,
$\text{Y}_B = (8.55 - 8.77) \times 10^{-11}$, which has recently been reported by Planck 2015 \cite{Ade:2015xua}.

 The novel Yukawa-coupling texture of our model leads to the normal neutrino mass hierarchy with $m_1 =0$ and a broken tri-bimaximal neutrino mixing pattern, as a result of which there is an automatic generation of nonvanishing reactor mixing angle. It is interesting to note that the model under consideration can accommodate non zero reactor mixing angle along with the explanation for non zero BAU without the help of any separate perturbation. From the necessity of see-saw to work a number of right handed neutrinos are introduced which are Majorana by nature. These extra right handed neutrinos can address $\text{Sin}^2\theta_{13} \neq 0$ by modifying the light neutrino mass matrix in such a way and at the same time offer explanation for the tiny excess of matter over antimatter via the process of leptogenesis. The complex Dirac Yukawa couplings of this system give rise to a nonzero lepton asymmetry which in turn yields the observed BAU. We have studied the lepton asymmetry generated over a range of right handed neutrino mass thereafter exploring the possibility of having both flavored and unflavored leptogenesis. Several BSM frameworks \citep{Borah:2013bza,Borah:2015vra,Kalita:2014vxa,Kalita:2014mga} are available in the literature where baryon asymmetry is produced by thermal leptogenesis with hierarchical right-handed neutrinos. These heavy Majorana neutrinos decay to SM particles violating lepton number, which later on gets converted into baryon number by non-perturbative sphaleron interaction. The goal of this paper is to study the cosmological baryon number asymmetry produced as a consequence of the presence of the fourth($\text{N}_4$) and fifth generation of RHN ($\text{N}_5$), the presence of which also introduces a non-vanishing reactor mixing angle in the theory. Thus, we aim here at presenting the BAU in terms of the set of light neutrino model parameters that gives rise to correct neutrino data. In this piece of work we have kept our analysis only upto finding the matter-antimatter asymmetry produced via the mechanism of leptogenesis. Two different kinds of vev alignments for the extra flavon $\eta$ are chosen in order to thoroughly study the affect of the same on observed neutrino parameters and BAU. Although the two different vev alignments lead to two different kinds of Dirac neutrino mass structures, the final light neutrino mass matrix remains the same. This fact permits us to keep the study for neutrino phenomenology as same as it is there in the original work by \cite{Meloni:2010sk}. However these two different Dirac mass matrices bring a little modification in lepton asymmetry calculation, which we discuss in numerical analysis section. 

This work is solely dedicated to explore the possibility of foreseeing baryogenesis via leptogenesis through the realization of a broken $\mu-\tau$ symmetric mass matrix. Rest of the paper has been planned in the following manner, in Section \ref{sec2} we present the model. Section \ref{sec3} is discussed with type I seesaw and Leptogenesis. Section \ref{sec4} is kept for numerical analysis and results. Finally in Section \ref{sec5} we end up with conclusion.

\section{The Model}
\label{sec2}
Among the variant non-Abelian discrete flavor symmetry groups $\text{A}_4$ stands out as the most appealing group in the context of understanding neutrino mass and mixing properties. It has shown a promising role in explaining the origin of tri-bi-maximal type of neutrino mixing since long. Keeping this in mind, this symmetry group has been chosen to explain neutrino mass and mixing. Although $A_4$ merely needs an introduction, still we slightly describe some properties of this group and how have they been utilized to give structures to the Dirac and Majorana mass matrices of the model. The non-Abelian group $A_4$ is the first alternating
group and is isomorphic to the tetrahedral group $\text{T}_d$. $\text{A}_4$ has four irreducible representations, among them there are three singlets $1, 1^{\prime},1^{\prime \prime}$ and one triplet 3. The group $\text{A}_4$ has 12 elements, which can be written in terms of the generators of the group
S and T. Where the generators satisfy the following relation (for detail one may refer to \cite{discreteRev})
\begin{equation*}
S^2 = (ST)^3= (T)^3 = 1
\end{equation*}

We consider the model discussed in \cite{Meloni:2010sk} for the purpose of studying baryogenesis via leptogenesis through the CP violating decay of the lightest RHN present in the model. In this model we have a total of five right handed neutrinos, among them three are the components of $\text{N}_T$ which transform as an $\text{A}_4$ triplet and hence are degenerate. The other two viz., $\text{N}_4$ and $\text{N}_5$ transform as $A_4$ singlets $1^\prime$ and $1^{\prime \prime}$ respectively. Therefore the SM fermion sector has been extended by the inclusion of three $SU(2)$ fermion singlets. At the same time there is an extra flavon $\eta$ which is kept as an $SU(2)$ doublet and $\text{A}_4$ triplet. The full particle content of the model has been shown in table \ref{tab2}.      
\begin{table}[htb]
\begin{tabular}{|c|c|c|c|c|c|c|c|c|c|c|c|}
\hline  & $ L_{e} $ & $L_\mu$ & $L_\tau$ & $ l_e^c $ & $l_\mu^c$  & $l_\tau^c$ & $N_T$ & $N_4$ & $N_5$ & $H$ & $\eta$ \\
\hline $SU(2)$ & 2 & 2 & 2 & 1 & 1 & 1 & 1 & 1 & 1 & 2 &2 \\
$A_4$ & 1 & $1^{\prime}$ & $1^{\prime \prime}$ & 1 & $1^{\prime \prime}$ & $1^{\prime }$ & 3 & $1^{\prime}$ & $1^{\prime \prime}$ & 1 &3 \\
\hline     
\end{tabular}
\caption{Fields and their transformation properties under $ SU(2) $ and $ A_{4} $ symmetry} \label{tab2}
  \end{table} 
The Yukawa Lagrangian for the neutrino sector can be written as
\begin{gather}
\begin{split}
\mathcal{L}_Y & = Y_1^{\nu} L_e(N_T \eta)_1 + Y_2^{\nu} L_\mu(N_T \eta)_1^{\prime \prime}+ Y_3^{\nu} L_\tau(N_T \eta)_1^{\prime}+ Y_4^{\nu} L_\tau N_4H +Y_5^{\nu} L_\mu N_5H \\ 
&  + M_1 N_T N_T + M_2 N_4 N_5 + h.c. \nonumber
\end{split}
\end{gather}
Following the $\text{A}_4$ product rules as mentioned in Appen.\ref{appen1} we can arrive at the following structures for Dirac mass matrix by considering two different vev alignments e.g. when $\eta$ takes vev as $\langle \eta \rangle \sim v_{\eta}(1,0,0)$ and $\langle \eta \rangle \sim v_{\eta}(1,1,1)$  respectively .
 \begin{equation}\label{eq:dirac1}
      m_{D1} = \left(\begin{array}{ccccc}
      y_1^\nu v_\eta & 0 & 0 & 0 & 0\\
      y_2^\nu v_\eta & 0 & 0 & 0 & y_5^\nu v_h\\ 
      y_3^\nu v_\eta & 0 & 0 & y_4^\nu v_h & 0
      \end{array}\right),  m_{D2} = \left(\begin{array}{ccccc}
      y_1^\nu v_\eta & y_1^\nu v_\eta & y_1^\nu v_\eta & 0 & 0\\
      y_2^\nu v_\eta & y_2^\nu v_\eta & y_2^\nu v_\eta & 0 & y_5^\nu v_h\\ 
      y_3^\nu v_\eta & y_3^\nu v_\eta & y_3^\nu v_\eta & y_4^\nu v_h & 0
      \end{array}\right)
   \end{equation}
   
 
In the same way we obtain the following structure for Majorana neutrino mass,
 \begin{equation}\label{eq:majorana}
      M_R = \left(\begin{array}{ccccc}
      M_1 & 0 & 0 & 0 & 0\\
      0 & M_1 & 0 & 0 & 0\\ 
      0 & 0 & M_1 & 0 & 0 \\
      0 & 0 & 0 & 0 & M_2 \\
      0 & 0 & 0 & M_2 & 0 
      \end{array}\right).
   \end{equation}
Irrespective of the two different Dirac mass matrix the complete light neutrino mass matrix structure remains almost the same. Therefore, neutrino parameters continue to be similar for the later case. Since the lepton asymmetry parameter depends on the complex Dirac Yukawa couplings as seen from the Eq. (\ref{eq:asymmetry}) we study the variation brought out by these two structures by considering two different vev alignments of the the $SU(2)$ doublet field $\eta$.      
\section{Type I seesaw mechanism and its consequence Leptogenesis}
\label{sec3}
 Having set the stage we can write the effective light neutrino mass matrix from the type I seesaw realization.
 \begin{equation}\label{typeI}
 -m_{\nu}\sim m_D^T M_R^{-1}m_D
\end{equation}   
The Light neutrino mass matrices we obtain with the help of Eq.(\ref{eq:dirac1}) and Eq.(\ref{eq:majorana}) are as follows
\begin{equation}\label{eq:neutrinomass}
 m_{\nu 1} = \left(\begin{array}{ccc}
      a^2 & ab & ac \\
      ab & b^2 & bc+k \\ 
      ac & bc+k & c^2
      \end{array}\right), m_{\nu 2} = \left(\begin{array}{ccc}
      3a^2 & 3ab & 3ac \\
      3ab & 3b^2 & 3bc+k \\ 
      3ac & 3bc+k & 3c^2
      \end{array}\right)
\end{equation}
where, the elements are defined as $a = \frac{y_1^\nu v_\eta}{\sqrt{M_1}} , b =  \frac{y_2^\nu v_\eta}{\sqrt{M_1}} , c =  \frac{y_3^\nu v_\eta}{\sqrt{M_1}}, k =  \frac{y_4^\nu y_5^\nu v_h^2}{M_2}$. 
 One of the consequences of the type I seesaw is the lepton asymmetry produced by the decay of the lightest RHN present in the system.
\subsection{Leptogenesis}
Since as a crucial requirement of fulfilling the seesaw mechanism, RHNs are already present in the model, this fact gives us the opportunity to study leptogenesis scenario in the model under consideration. Lepton asymmetry is created by the decay of the lightest RHN present in the model. It is to note that all the Dirac Yukawa couplings coming from the type I seesaw are complex and hence can act as a source of CP-violation, as there are no CP-violating phase associated with RHNs. Therefore it can be said that the lone contribution to CP-asymmetry is coming from the complex Dirac Yukawa couplings in the present context.

\begin{figure*}[h!]
\begin{center}
\includegraphics[width=15cm,height=3cm]{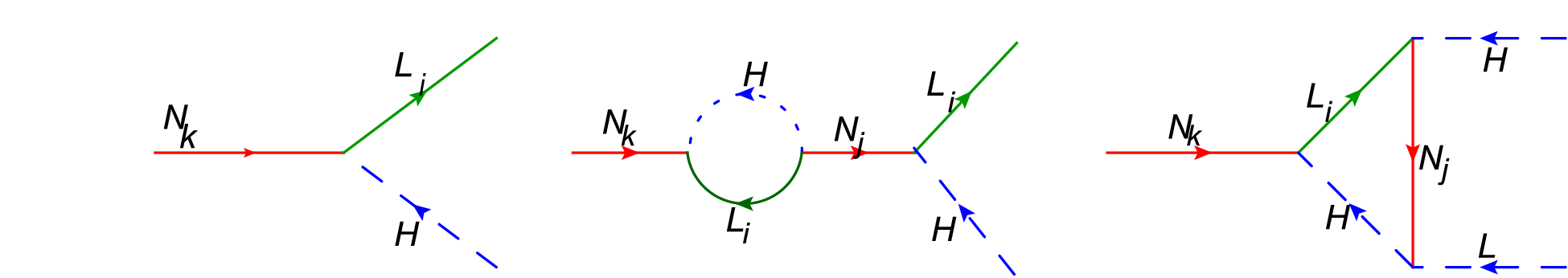}
\caption {Feynman diagram contributing to L violating RHN decay}
\label{fig1}
\end{center}
\end{figure*}

 \begin{gather}\label{eq:asymmetry}
 \begin{align}
    \epsilon_{1}^\alpha = \frac{1}{8\pi v^2} \frac{1}{(m_{LR}^{\dagger} m_{LR})_{11}}\sum_{j= 2,3}\text{Im}[(m_{LR}^{*})_{\alpha 1}(m_{LR}^{\dagger} m_{LR})_{1j}(m_{LR})_{\alpha j}]g(\text{x}_j)\\ \nonumber
 + \frac{1}{8\pi v^2}\frac{1}{(m_{LR}^{\dagger} m_{LR})_{11}}\sum_{j= 2,3}\text{Im}[(m_{LR}^{*})_{\alpha 1}(m_{LR}^{\dagger} m_{LR})_{j1}(m_{LR})_{\alpha j}]\frac{1}{1-\text{x}_j}
  \end{align}
 \end{gather}    
 with $v = 174$ GeV, the vev of the SM higgs doublet.
 \begin{equation*}
 g(\text{x})= \sqrt{\text{x}}\Bigg(1+\frac{1}{1-\text{x}}-(1+\text{x})\text{ln}\frac{1+\text{x}}{\text{x}}\Bigg) 
 \end{equation*}  
with $\text{x}_j = \frac{M_j^2}{M_{1}^2}$. When we sum over all the flavors $\alpha = e, \mu, \tau$ popularly which is known as one flavor or unflavored leptogenesis, the second term in the expression vanishes. The sum over all flavors then can be written as
\begin{equation}\label{eq:unflavored}
\epsilon_1 = \frac{1}{8\pi v^2} \frac{1}{(m_{LR}^{\dagger} m_{LR})_{11}}\sum_{j= 2,3} \text{Im}[(m_{LR}^\dagger m_{LR})^2_{1j}]g(\text{x}_j)
\end{equation}    

From the expression above for lepton asymmetry, one can write the final BAU as,
\begin{equation}\label{eq:bau}
Y_B = c \kappa \frac{\epsilon}{g_*}
\end{equation}
where $c$ determines the fraction of lepton asymmetry being converted into baryon asymmetry, the value of $c$ is found to be -0.55. $\kappa$ is the dilution factor responsible for the wash out processes which erase the generated asymmetry. One can have following expressions for $\kappa$ (for detail you may see \cite{sphaleron,Pilaftsis:1998pd,Flanz:1998kr})depending on the scale of the wash out factor $K$.
\begin{gather}\label{eq:washout}
-\kappa \approx \sqrt{0.1K} \text{exp}[-4/(3(0.1K)^{0.25})],\hspace{6mm} \text{for} \hspace{6mm}    K \geq 10^6\\
\approx \frac{0.3}{K(\text{ln} K)^{0.6}}, \hspace{6mm} \text{for}  \hspace{6mm}        10 \leq K \leq 10^6\\
\approx \frac{1}{2\sqrt{K^2+9}},\hspace{6mm} \text{for}  \hspace{6mm}    0\leq K \leq 10.
\end{gather}
where, $K$ quantifies the deviation of the decay rate of the lightest RHN from the expansion rate of the universe and is parametrized as,
\begin{equation}
K = \frac{\Gamma_1}{H(T = M_1)}= \frac{(m_{LR}^{\dagger}m_{LR})_{11} M_1}{8 \pi v^2}\frac{M_{Pl}}{1.66 \sqrt{g_*}M_1^2}
\end{equation}
where, $\Gamma_1$ is the decay width of the lightest decaying RHN and H is the Hubble rate of expansion at temperature $T = M_1$.
We denote the effective number of relativistic degrees of freedom as $g_*$ and is approximately 110. It is worth mentioning that only the decay of the lightest RHN  serves effective for a given hierarchical RHN spectra, as the asymmetry produced by the decay of heavier RHNs gets washed out notably. Therefore, $\epsilon_1^\alpha$ is the only germane quantity which contributes to the total lepton asymmetry. There are three regimes of baryogenesis depending on the scale of the decaying right handed neutrino mass. The expression for lepton asymmetry as given in the Eq.(\ref{eq:asymmetry}) implies the asymmetry generated by summing over all the flavors. The expression given in the Eq.(\ref{eq:unflavored}) works in the regime when the decaying RHN mass goes greater than or equal to $10^{12}$ GeV, when all the flavors behave similarly and are out of equilibrium. Now if the mass of the decaying RHN falls with the range $10^9 < \text{M}_{N1}<10^{12}$ GeV then, the $\tau$ flavor is in equilibrium and distinguishable and the scenario is familiar as tau-flavor or two flavored leptogenesis. In this domain there are two relevant lepton asymmetry parameters $\epsilon_1^e$ and $\epsilon_1^\mu$. One can write the final BAU for this regime as given by $\text{Y}_B^{\text{2flavor}}$ in Eq.(\ref{flavored2}). Again if the RHN mass falls less than $10^9$ GeV , it is termed as three flavored or fully flavored leptogenesis, where all the flavors are in equilibrium and hence distinguishable. For this mass scale of the RHN the final BAU is estimated by an expression given in Eq.(\ref{flavored3}).

\begin{equation}\label{flavored2}
Y_B^{\text{2flavor}}= -\frac{12}{37g_*} [ \epsilon_2 \eta \Big(\frac{417}{589} \tilde{m_2}\Big)+\epsilon_1^\tau \eta \Big(\frac{390}{589}\tilde{m_\tau}\Big)]
\end{equation}
\begin{equation}\label{flavored3}
Y_B^{\text{3flavor}}= -\frac{12}{37g_*} [ \epsilon_1^e \eta \Big(\frac{151}{179} \tilde{m_e}\Big)+\epsilon_1^\mu \eta \Big(\frac{344}{537}\tilde{m_\mu}\Big)+ \epsilon_1^\tau \eta \Big(\frac{344}{537}\tilde{m_\tau}\Big)]
\end{equation}
where, $\epsilon_2 = \epsilon_1^e + \epsilon_1^\mu, \tilde{m_2} = \tilde{m_e} +\tilde{m_\mu}, \tilde{m_{\alpha}}= \frac{(m_{LR}^*)_{\alpha 1} (m_{LR})_{\alpha 1}}{M_1}$. One can write the expression for $\eta$ as
\begin{equation*}
\eta(\tilde{m_{\alpha}})= \Bigg[\Big(\frac{\tilde{m_{\alpha}}}{8.25 \times 10^{-3} \text{eV}}\Big)^{-1} + \Big(\frac{0.2 \times 10^{-3} \text{eV}}{\tilde{m_{\alpha}}}\Big)^{-1.16}\Bigg]^{-1}
\end{equation*}

For numerical analysis we have diagonalised the right handed neutrino mass matrix for getting the eigenvalues of $M_{R}$ that we show in numerical analysis section. The Dirac mass matrix has been chosen in a basis where RH neutrino mass matrix is diagonal. For that one can write $m_{LR} = m_D U_R$, where $U_R^*M_{RR}U_R^\dagger = \text{diag}(M_1, M_2,M_3,M_4,M_5)$.

\section{ Numerical Analysis and results}
\label{sec4}
The light neutrino mass matrix that we obtain with the help of type I seesaw formula using Eq.(\ref{eq:dirac1}) and Eq.(\ref{eq:majorana}) yields a $\mu -\tau$ symmetry broken structure which automatically takse non-zero reactor mixing angle into account. The two different vev alignments  gives rise to two different Dirac mass matrix and hence two different light neutrino mass matrices. This fact allows us to study leptogenesis for two Dirac Yukawa coupling matrices. From Eq.(\ref{eq:neutrinomass}) it is clear that however we have two seperate $m_D$ to feed into the type I seesaw formula, the final light neutrino mass matrices obtained for each case are almost similar and yields the neutrino phenomenology in relatively same manner. The two Dirac mass matrices($m_{D1}$, $m_{D2}$) are different from each other in a sense that, $m_{D2}$ has more number of non-zero entries in compaison to $m_{D1}$, which brings modification in the leptogenesis calculation as the CP asymmetry is dependent on the Dirac Yukawa coupling as seen from the Eq.(\ref{eq:asymmetry}). Apart from studying leptogenesis, the first kind of vev alignment of the flavon $\eta$ allow us to declare it as a potential dark matter candidate, as shown in \cite{Meloni:2010sk, Hirsch:2010ru, Mukherjee:2015axj}. It is interesting to note that the presence of the 5th RHN $N_5$ results into the light neutrino mass matrix structure in a way that the $\mu-\tau$ symmetry gets broken yielding $\theta_{13}$ with a nonzero value which falls within the $3\sigma$ range. Authors in \cite{Meloni:2010sk} has explained the reason of why only normal hierarchy is supported by the presented model. Thus we have calculated the lepton asymmetry only for the parameter space where neutrino mass follow normal hierarchy. There are five eigenvalues for the Majorana mass matrix $M_R$, among which three belong to the same $A_4$ triplet and hence exactly degenerate. Therefore diagonalization of $M_R$ gives rise to two distinct hierarchical eigenvalues of the RH neutrino mass matrix. Among them we may chose any one to be slightly lighter than the other, CP-violating decay of which to SM leptons and Higgs is supposed to create lepton asymmetry.

Since both kinds of vev alignment leads to almost similar kind of light neutrino mass matrix, we have chosen the same way for calculating the light neutrino model parameters for each case($m_{D1}$ and $m_{D2}$). It is known that a complex symmetric matrix has got 12 independent real parameters, among which there are three unphysical phases after readsorbing whom the number of parameter comes down to nine. In this work the light neutrino mass matrix Eq. (\ref{eq:neutrinomass}) has four complex parameters and hence five real independent parameters eg., $|a|,|b|,|c|$  and the modulous and phase of the combination $bc+k = d e^{i \phi_d}$ as derived in \cite{Meloni:2010sk}. With the new phase relationship the light neutrino mass matrix one can write in the following form
\begin{equation}
 m_{\nu 1} = \left(\begin{array}{ccc}
      a^2 & ab & ac \\
      ab & b^2 & d e^{i \phi_d} \\ 
      ac & d e^{i \phi_d} & c^2
      \end{array}\right)
\end{equation}
Now we can make three equations in order to relate three parameters e.g. a, b and c with the light neutrino masses using the following set of equations.   
\begin{gather}\label{eq:relate}
\text{Tr}(m^2_\nu) = \text{t} = a^4+2a^2(b^2+c^2)+b^4+c^4+2d^2 = m_{\nu 1}^2+m_{\nu 2}^2+m_{\nu 3}^2 \nonumber\\
 \text{det}(m_\nu ^2)= a^4(b^2 c^2 - 2bcd cos\phi_d + d^2)^2 = m_{\nu 1}^2 m_{\nu 2}^2 m_{\nu 3}^2 \\ 
 \frac{1}{2}\big[t^2 - \text{Tr}(m_{\nu}^2 m_{\nu}^2)\big] = (b^2 c^2 - 2bcd cos \phi_d +d^2) 
 \times \big[2a^4 +2a^2(b^2+c^2)+b^2 c^2+ 2bcd cos \phi_d +d^2\big] \nonumber\\
 = m_{\nu 1}^2(m_{\nu 2}^2+m_{\nu 3}^2)+ m_{\nu 2}^2 m_{\nu 3}^2 \nonumber
\end{gather}
where, we denote "Tr" and "det" as trace and determinant of the light neutrino mass matrix. We did a random scan of some of the independent parameters in order to find a, b and c. From the above relations we see that a, b and c can be written in terms of the light neutrino mass eigenvalues. Since this analysis is only restricted to normal hierarchy mass pattern thus we can have $m_1, m_2 = \sqrt{\Delta m_{\text{sol}}^2 + m_1^2}, m_3 = \sqrt{\Delta m_{\text{sol}}^2 + \Delta m_{\text{atm}}^2 + m_1^2}$ as the three light neutrino masses. We choose $m_1, \Delta m_{\text{sol}}^2, \Delta m_{\text{atm}}^2$ as the independent parameters and are randomly varied within their $3\sigma$ range as reported by \cite{deSalas:2017kay} for numerical computation. On the other hand among the model parameters $d$ and $\phi_d$ have been chosen for random scan within the interval mentioned below.

\begin{gather}\label{eq:data}
\Delta m_{\text{sol}}^2 = (7.05 - 8.14) \times 10^{-5} \text {eV}^2,\hspace{4mm}  \Delta m_{\text{atm}}^2 =(2.43 - 2.67) \times 10^{-3} \text {eV}^2 ,\\ 
  m_1 = (10^{-5} - 1) \text eV,\hspace{4mm} d \in [-1,1],\hspace{4mm} \phi_d \in [-\pi, \pi) \nonumber
\end{gather}
 As already mentioned that for the two vev alignments the light neutrino mass matrices are almost similar thus we determine the model parameters with the help of the above mentioned equations. With the values of a, b, c found from the solutions we construct the Dirac Yukawa coupling matrices for each case separately. Once the two different $m_D$s are constructed, with the help of them lepton asymmetry parameter is calculated for each $m_D$ type i.e., $m_{D1}$ and $m_{D2}$. To find the Dirac Yukawa couplings for each case following definitions are used during calculation.
\begin{gather*}
  a_1 = y_1^\nu v_\eta, a_2 = y_2^\nu v_\eta, a_3 = y_3^\nu v_\eta, a_4 = y_4^\nu v_h, a_5 = y_5^\nu v_h\\
  \text{VEVI case} :\frac{a_1^2}{f} = a^2  , \frac{a_2^2}{f} = b^2 , \frac{a_3^2}{f} = c^2 , \frac{a_4 a_5}{g} = k\\
 \text{VEVII case} : \frac{3 a_1^2}{f} = a^2 ,\frac{3 a_2^2}{f} = b^2, \frac{3 a_3^2}{f} = c^2
\end{gather*}
For numerical convenience we take $y_4 = y_5$ which results into $a_4 = a_5$. Using this approximation we evaluate $a_1 , a_2, a_3, a_4$. Using this set of solutions we construct $m_D$ for each case. 
 \begin{equation}\label{eq:dirac}
      m_{D1} = \left(\begin{array}{ccccc}
      a_1 & 0 & 0 & 0 & 0\\
      a_2 & 0 & 0 & 0 & a_5\\ 
      a_3 & 0 & 0 & a_4 & 0
      \end{array}\right),  m_{D2} = \left(\begin{array}{ccccc}
     a_1 & a_1 & a_1 & 0 & 0\\
      a_2 & a_2 & a_2 & 0 & a_5\\ 
      a_3 & a_3 & a_3 &a_4 & 0
      \end{array}\right)
   \end{equation}

As already mentioned in Sec.\ref{sec3} there are three regimes of leptogenesis, one flavor, two flavor and three flavor depending on the RHN mass scale. Now we may investigate whether the presented framework has some parameter space for each of the three regimes or not. For that, we choose the mass scales of RHN to be of three ranges. We choose the parameter g to be slightly larger than f in the majorana mass matrix by denoting f as $M_i$ and g as $M_j$. For one flavour regime we have chosen $M_i = 10^{12} $ GeV and $M_j= 10^{13}$ GeV which is the specified mass domain for unflavored leptogenesis as mentioned in \cite{Samanta:2017kce}. For tau-flavor leptogenesis the RHN mass has been chosen around $10^{10}$ GeV. As demanded by the corresponding mass regime for RHN, $M_i, M_j$ have been chosen less than $10^9$ GeV. We have calculated the lepton asymmetry $\epsilon_i^\alpha$ by changing the ratio of RHN mass squared $\frac{M_j^2}{M_i^2}$ for each regime of leptogenesis. We choose the RHN masses as required for different region of leptogenesis as shown in table \ref{tab3}.
\begin{table}[htb]
\begin{tabular}{|c|c|c|c|}
\hline & $M_i$ & $M_j$ & $ x_j= \frac{M_j^2}{M_i^2}$\\
\hline One flavor & $8 \times 10^{12}$ & $10^{13}$ & 1.5625   \\
\hline Two flavor & $10^{10}$ & $1.0009\times10^{10}$ & 1.0018    \\
\hline Three flavor & $10^{8}$ & $1.000009 \times 10^{8}$ & 1.00002  \\
\hline   
\end{tabular}
\caption{RHN masses and their mass squared ratios for different domain of leptogenesis} \label{tab3}
  \end{table} 

To start with, we first evaluate the model parameters a, b and c by  randomly varying the mass splittings in their allowed $3\sigma$ ranges along with the lightest mass as mentioned in Eq.(\ref{eq:data}). At the same time we have also varied $d$ and $\phi_d$ in their above mentioned specified ranges. With the help of the values found for a, b and c we found out the Dirac Yukawa couplings by considering three different ranges for heavy RHNs. With the help of the Dirac Yukawa couplings the Dirac mass matrix is constructed and with the help of that the lepton asymmetry for the leptonic sector is calculated. In order to find BAU using that value for lepton asymmetry. We have restricted our analysis for leptogenesis only upto normal hierarchy pattern of light neutrino mass as the model disfavor the IH mass pattern.

\subsection{VEV1 case}
\begin{figure*}
\begin{center}
\includegraphics[width=0.45\textwidth]{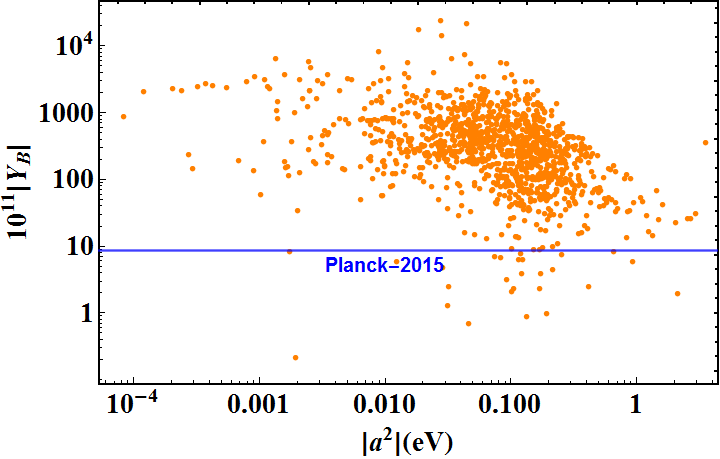}
\includegraphics[width=0.45\textwidth]{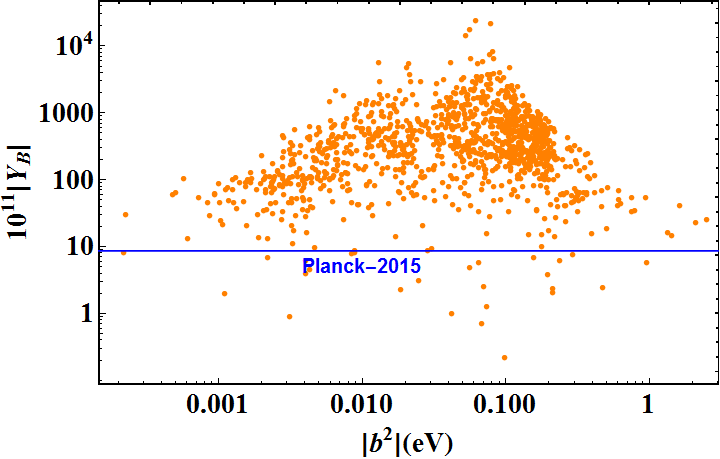} \\
\includegraphics[width=0.45\textwidth]{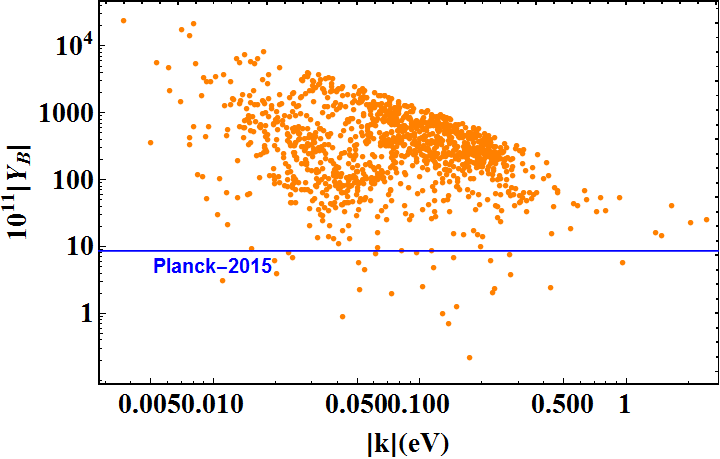}
\end{center}
\begin{center}
\caption{Dependence of BAU on various model parameters in case of unflavored leptogenesis with $M_N \geq 10^{12}$ GeV. The blue horizontal band represents the Planck bound for $\text{Y}_B = (8.55 - 8.77) \times 10^{-11}$.}
\label{fig2}
\end{center}
\end{figure*}

\begin{figure*}
\begin{center}
\includegraphics[width=0.45\textwidth]{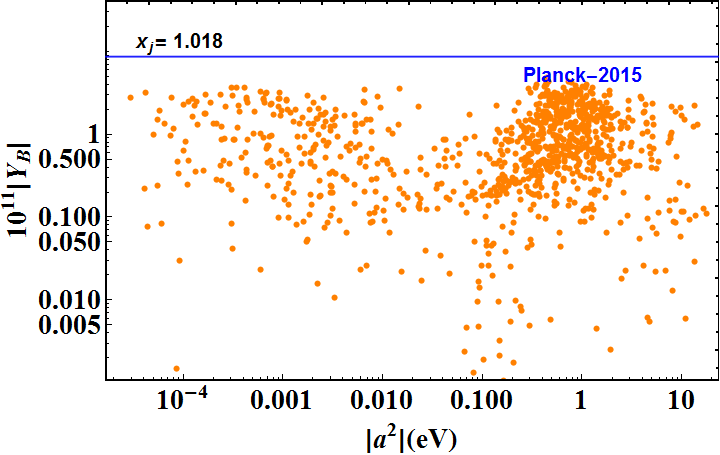}
\includegraphics[width=0.45\textwidth]{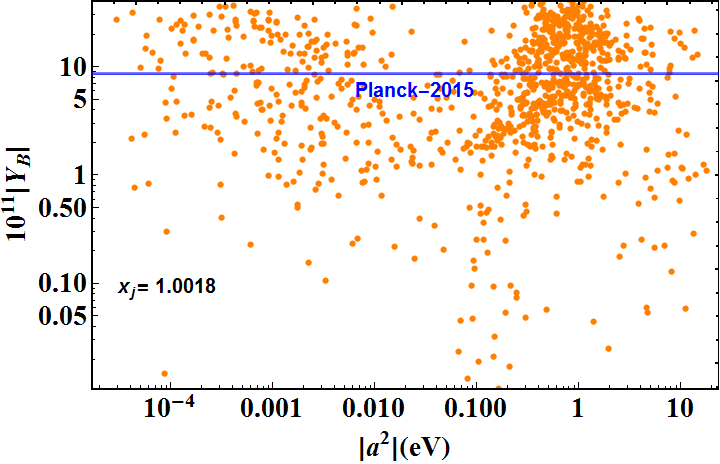} 
\includegraphics[width=0.45\textwidth]{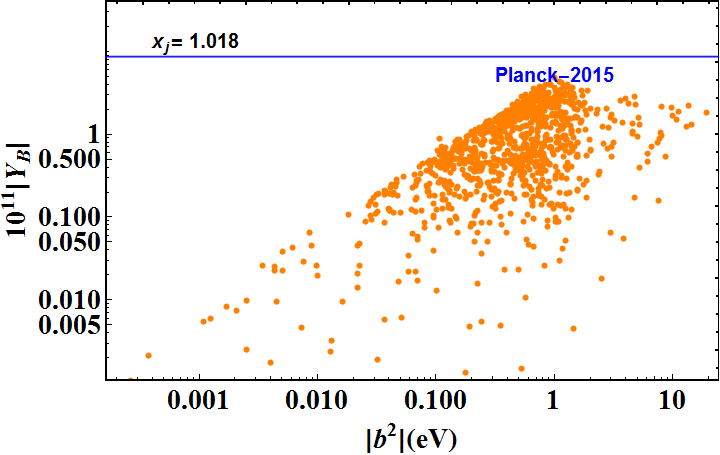}
\includegraphics[width=0.45\textwidth]{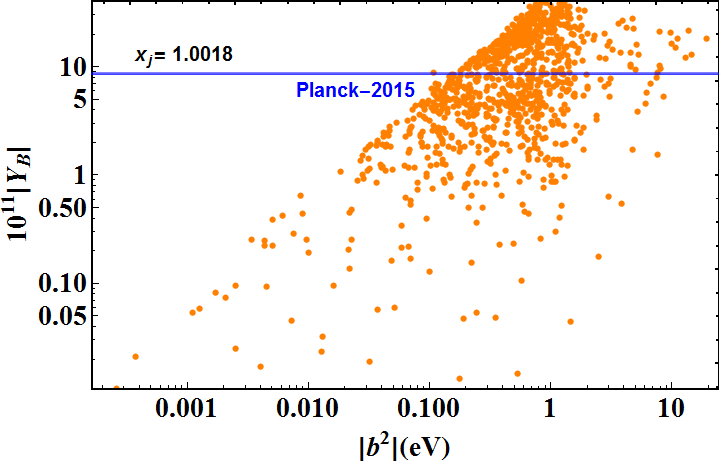}
\includegraphics[width=0.45\textwidth]{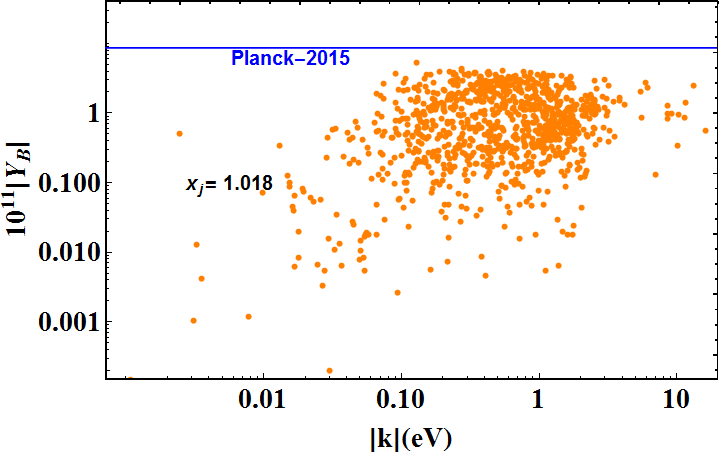}
\includegraphics[width=0.45\textwidth]{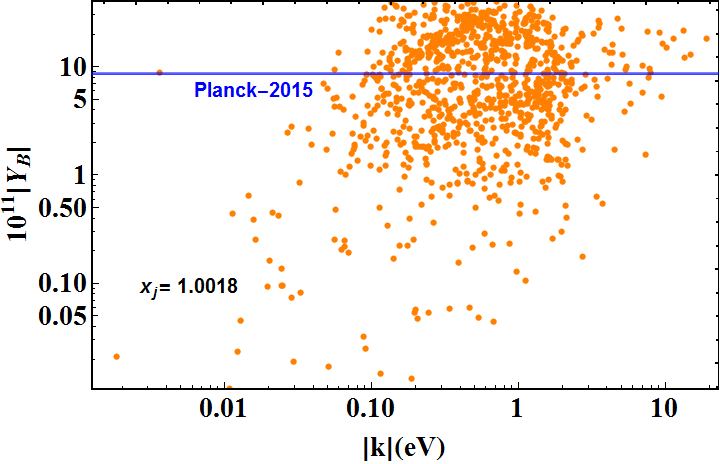}
\end{center}
\begin{center}
\caption{BAU versus model parameters for two flavored leptogenesis, where $10^{9} \text GeV < \text{M}_\text{N} < 10^{12}$ GeV. The left panel is kept for the mass squared ratio $\text{x}_j = \frac{M_j^2}{M_i^2} = 1.018$, whereas in the right one  $\text{x}_j = \frac{M_j^2}{M_i^2} = 1.0018$. The blue horizontal band represents the Planck bound for $\text{Y}_B = (8.55 - 8.77) \times 10^{-11}$.} 
\label{fig3}
\end{center}
\end{figure*}

\begin{figure*}
\begin{center}
\includegraphics[width=0.45\textwidth]{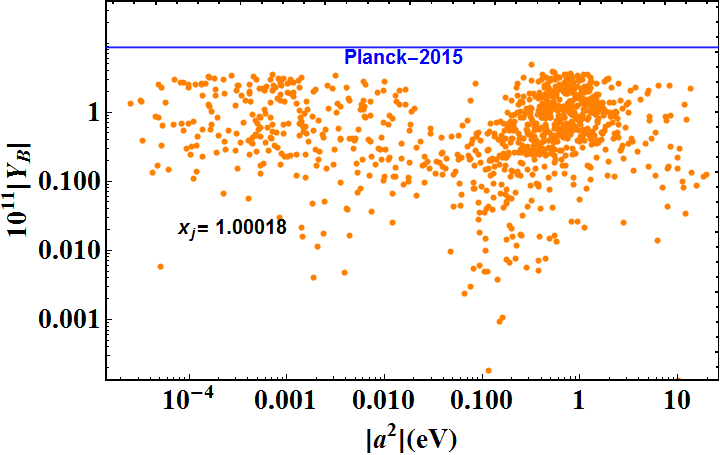} 
\includegraphics[width=0.45\textwidth]{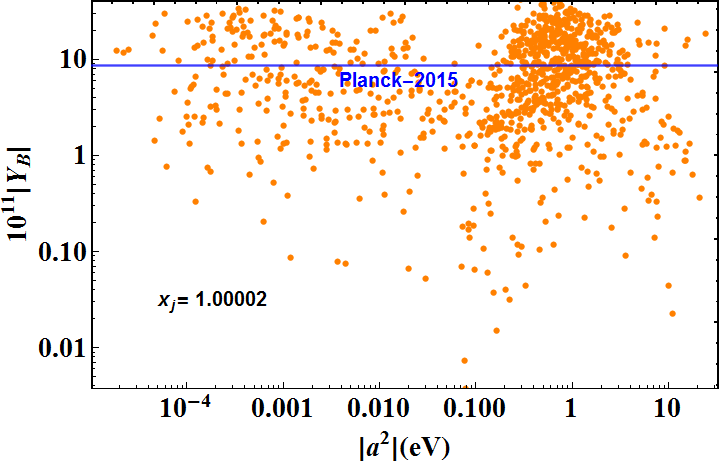} 
\includegraphics[width=0.45\textwidth]{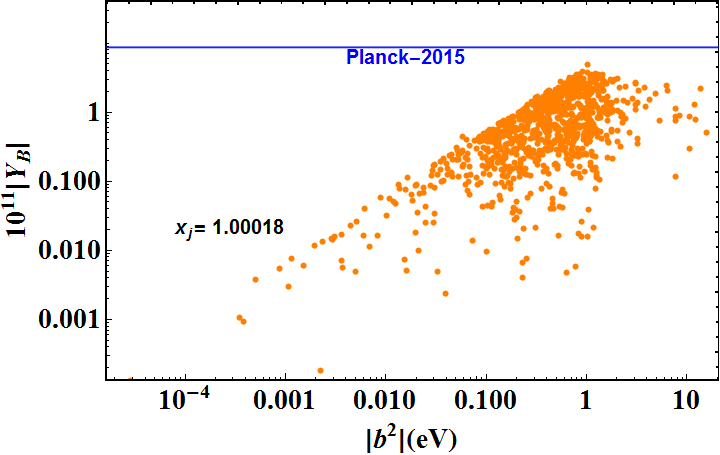}
\includegraphics[width=0.45\textwidth]{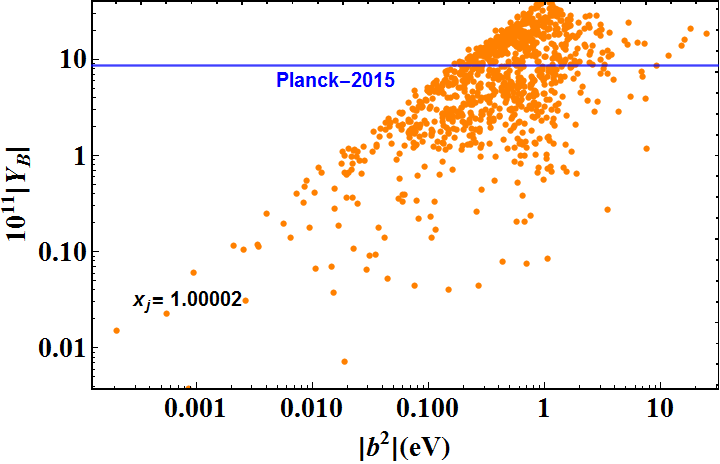}
\includegraphics[width=0.45\textwidth]{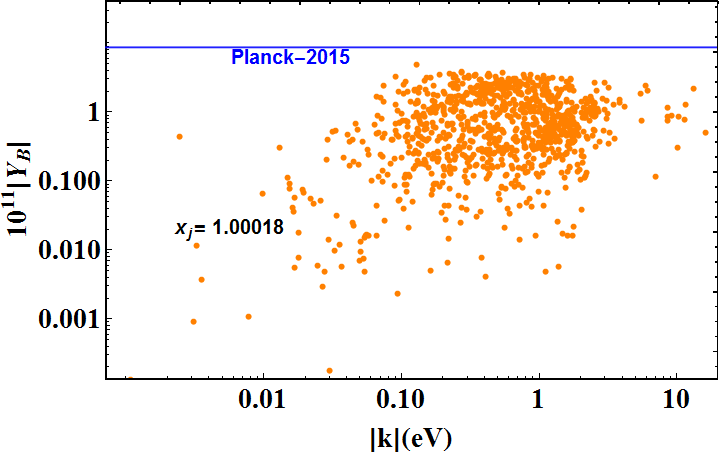}
\includegraphics[width=0.45\textwidth]{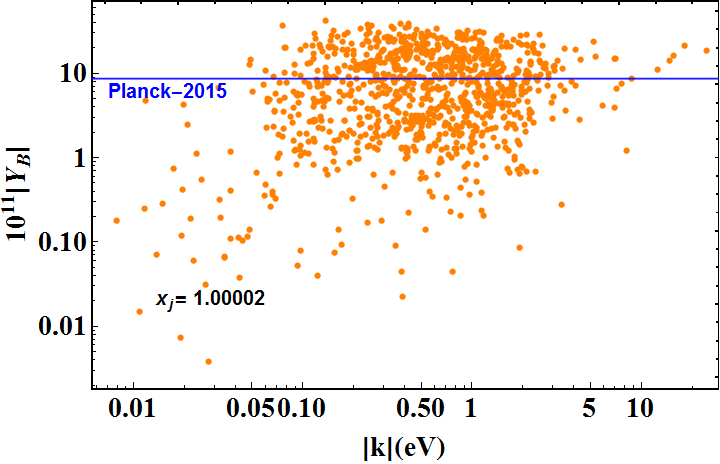}
\end{center}
\begin{center}
\caption{Dependence of BAU as a function of the model parameters for three flavored regime scenario with the lightest RHN mass $M_N \leq 10^{9}$ GeV. The blue horizontal band represents the Planck bound for $\text{Y}_B = (8.55 - 8.77) \times 10^{-11}$.}
\label{fig4}
\end{center}
\end{figure*}

\begin{figure*}
\begin{center}
\includegraphics[width=0.45\textwidth]{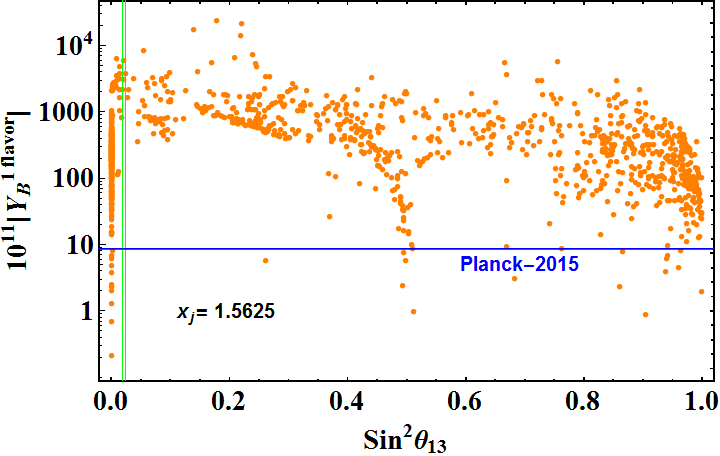}
\includegraphics[width=0.45\textwidth]{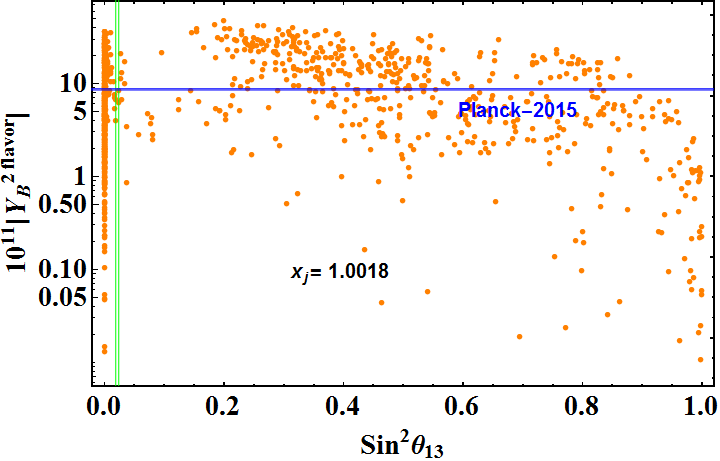}
\includegraphics[width=0.45\textwidth]{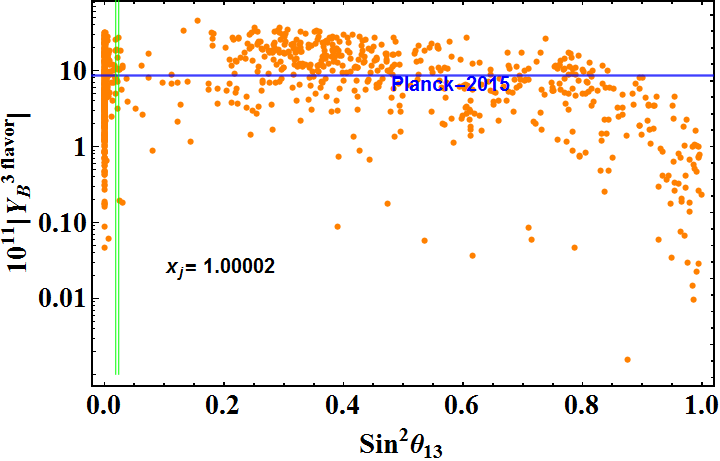}
\end{center}
\begin{center}
\caption{Parameter space for BAU and non zero reactor angle in case of one flavor(left of upper panel), two flavor(right of upper panel) and three flavor(lower panel) leptogenesis. The green vertical band presents the latest experimental bound for $\text{Sin}^2\theta_{13}$ with the blue horizontal band for the Planck bound for $\text{Y}_B = (8.55 - 8.77) \times 10^{-11}$.}
\label{fig5}
\end{center}
\end{figure*}

As already mentioned in the previous section we have computed the lepton asymmetry parameter for two type of Dirac mass matrices. Using that value of $\epsilon_i^\alpha$ baryon asymmetry is determined. We categorize the corresponding results for each Dirac mass matrices and presented them in separate subsections. In this subsection we show the results for the $m_D$ which results for the VEV alignment of $\eta$ as $v_\eta (1,0,0)$. We present here the BAU as a function of the model parameters. In Fig \ref{fig2} we plot for BAU as a function of the model parameters for the case of unflavored leptogenesis. BAU for the two flavor regime has been shown in Fig \ref{fig3}. In this plot we have shown the BAU for two different values of the Majorana mass splitting $\text{x}_j$, one for $\text{x}_j = 1.018$ and another for $\text{x}_j = 1.0018$. And from the two plots it is clear that, as the mass squared ratio decreases we can have more parameter space for baryon asymmetry of the universe for the chosen range of RHN mass. Fig \ref{fig4} evinces the BAU as a function of the model parameters for three flavored leptogenesis i.e., when all the flavors are out of equilibrium. Here also we have kept two values of $\text{x}_j$ and calculated the corresponding BAU for the chosen range of right handed neutrino mass pertinent to three flavored region of leptogenesis. From two and three flavor leptogenesis calculation one significant point is to be noted, that is as we go far below $10^{12}$GeV for the concerned RHN mass, the mass squared ratio $\text{x}_j$ shall go on decreasing to produce the required amount of CP asymmetry which accounts for the observed BAU. For unflavored leptogenesis scenario the ratio can be bigger compared to that required for flavored leptogenesis as the RHN mass falls larger than or equal to $10^{12}$GeV which is quite high. But as we go towards the lower mass regime for RHN the splitting demands a smaller value in order to give rise to adequate CP asymmetry to yield the observed baryon asymmetry. Fig \ref{fig5} presents the parameter space for $\text{Y}_B$ and $\text{Sin}^2 \theta_{13}$. If we look for a common parameter space for $\text{Y}_B$ and $\text{Sin}^2\theta_{13}$, a conclusion can be drawn from Fig
\ref{fig5} that, the model prediction for the same is much better for two and three flavor leptogenesis scenarios, while in the case of unflavored leptogenesis the plank bound  for the observed BAU does not meet the recent bound for non-zero reactor mixing angle. 

\clearpage
\clearpage

\subsection{VEV2 case}
\begin{figure*}
\begin{center}
\includegraphics[width=0.45\textwidth]{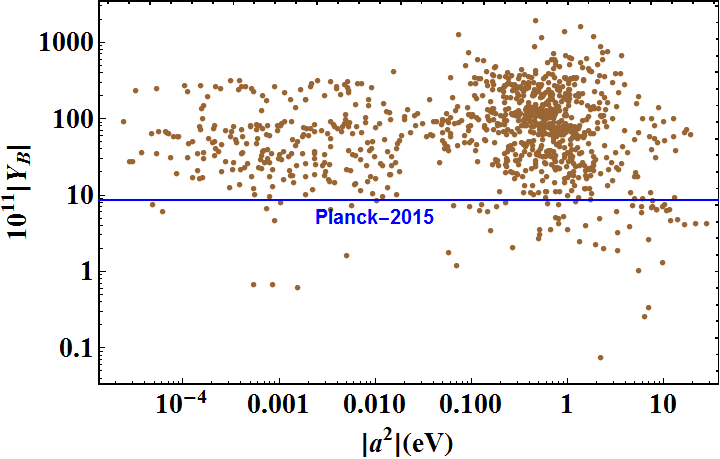}
\includegraphics[width=0.45\textwidth]{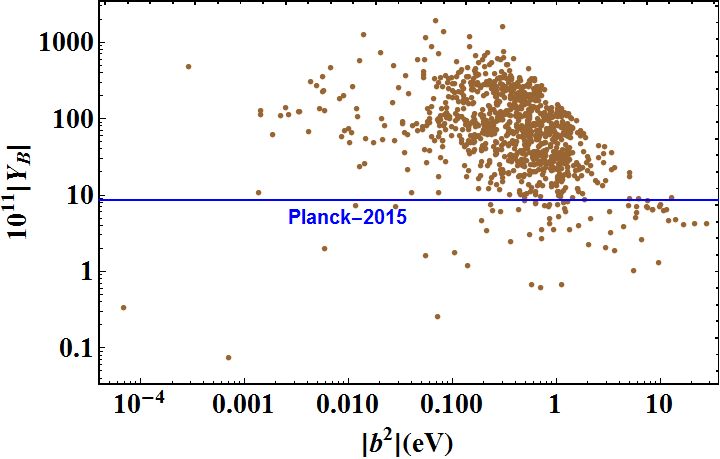} \\
\includegraphics[width=0.45\textwidth]{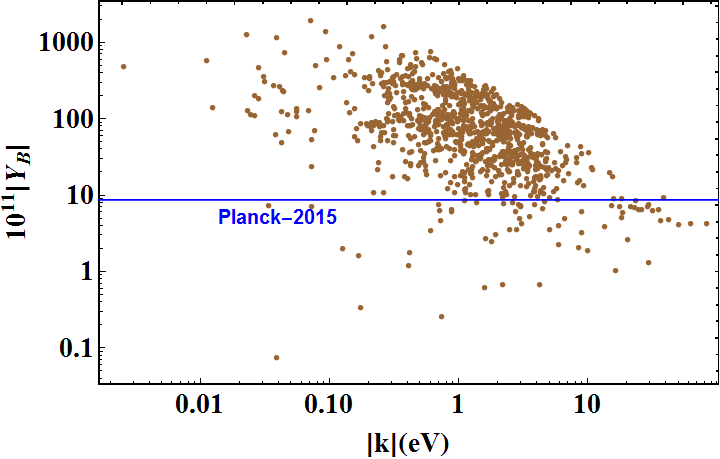}
\end{center}
\begin{center}
\caption{BAU as a function of the model parameters for one flavor leptogenesis scenario where the blue horizontal band represents the Planck bound for $\text{Y}_B = (8.55 - 8.77) \times 10^{-11}$.}
\label{fig6}
\end{center}
\end{figure*}

\begin{figure*}
\begin{center}
\includegraphics[width=0.45\textwidth]{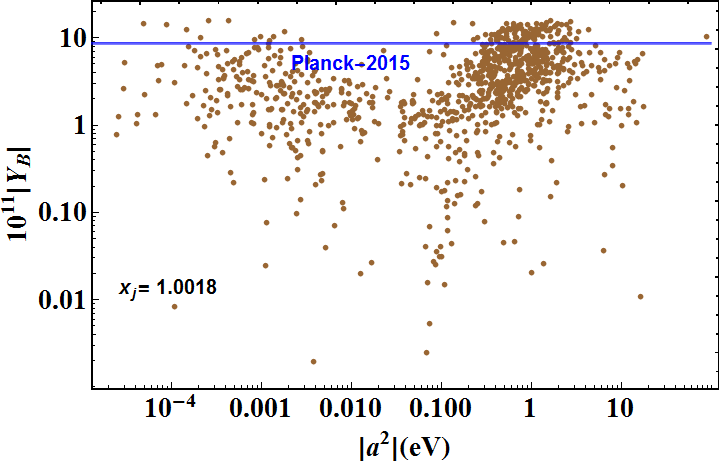}
\includegraphics[width=0.45\textwidth]{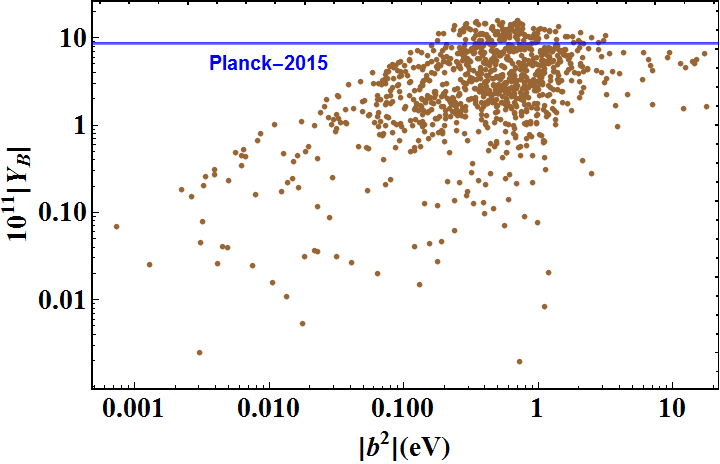} \\
\includegraphics[width=0.45\textwidth]{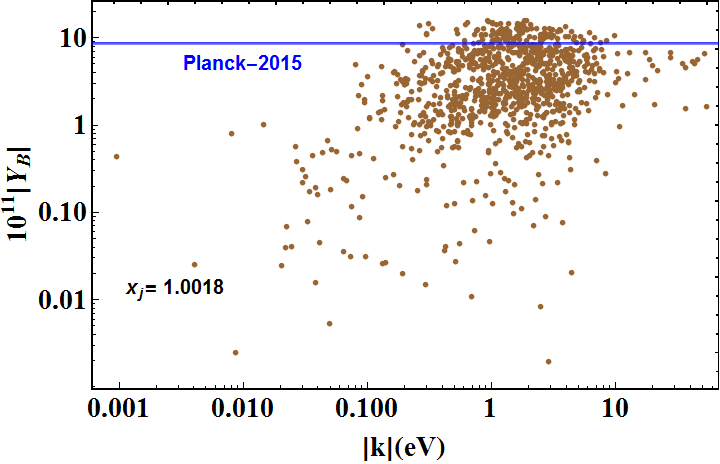}
\includegraphics[width=0.45\textwidth]{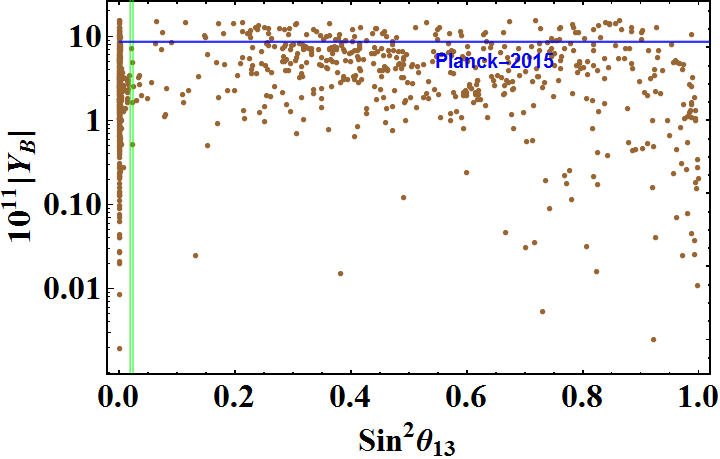}
\end{center}
\begin{center}
\caption{BAU as a function of the model parameters and nonzero reactor mixing angle for two flavored leptogenesis scenario with $\text{x}_j = 1.0018$. The green vertical band presents the latest experimental bound for $\text{Sin}^2\theta_{13}$ and the blue horizontal band shows the Planck bound for $\text{Y}_B = (8.55 - 8.77) \times 10^{-11}$.}
\label{fig7}
\end{center}
\end{figure*}

\begin{figure*}
\begin{center}
\includegraphics[width=0.45\textwidth]{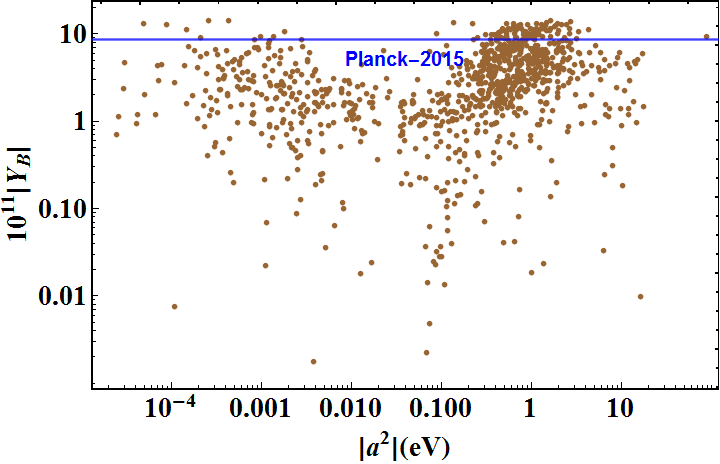}
\includegraphics[width=0.45\textwidth]{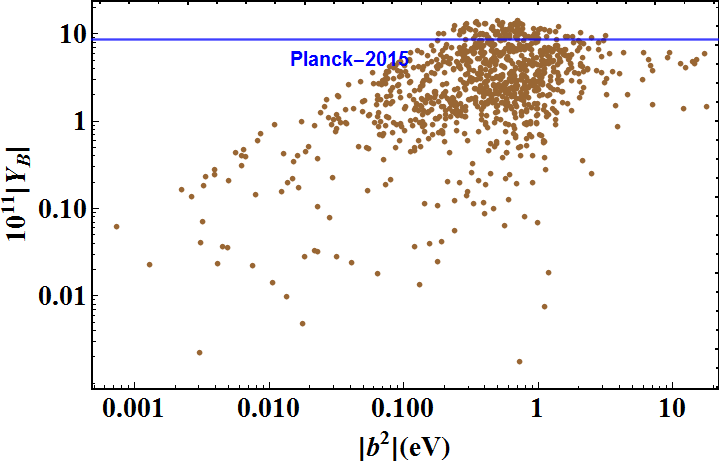} \\
\includegraphics[width=0.45\textwidth]{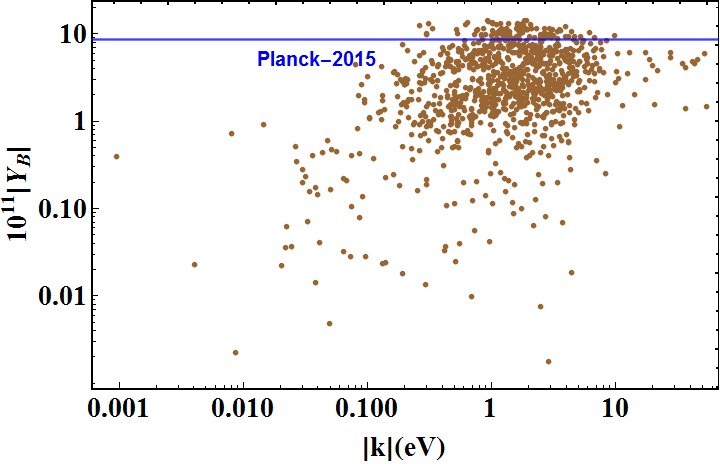}
\includegraphics[width=0.45\textwidth]{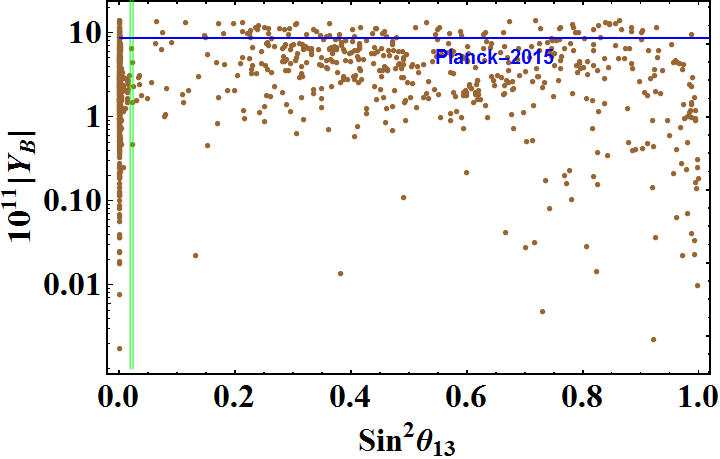}
\end{center}
\begin{center}
\caption{BAU as a function of the model parameters and nonzero reactor mixing angle for three flavored leptogenesis scenario with $\text{x}_j = 1.00002$. The green vertical band presents the latest experimental bound for $\text{Sin}^2\theta_{13}$ with the blue horizontal band showing the Planck bound for $\text{Y}_B = (8.55 - 8.77) \times 10^{-11}$.}
\label{fig8}
\end{center}
\end{figure*}

 We repeat the same procedure for determining BAU considering the Dirac mass matrix of the second kind. While computing lepton asymmetry for this case it is seen that unlike the previous case (with Dirac mass matrix of first kind), the Yukawa coupling associated with the fifth generation of RHN($\text{N}_5$) takes active part in producing a sufficient amount of lepton asymmetry required to generate the expected BAU. In Fig \ref{fig6} we present the BAU found with respect to the model parameters for unflavored leptogenesis scenario. For flavored leptogenesis we have chosen two possible mass domains of the decyaing RHN, one for two flavor regime and another for three flavor regime. Despite of this fact we need to chose two RHN masses($\text{M}_i$ and $\text{M}_j$) for each type of leptogenesis scenario(one, two and three flavor)in order to keep the mass squared ratio ($\text{x}_j = \frac{\text{M}_j^2}{\text{M}_i^2}$)needed for successful leptogenesis. Fig \ref{fig7} represents BAU as a function of model parameters and nonzero reactor mixing angle. In case of two/three flavored leptogenesis like the earlier case, for $\text{x}_j = 1.0018$ / $\text{x}_j = 1.00002$ we get the required lepton asymmetry which later on gets converted into baryon asymmetry by electroweak sphaleron. This mass splitting plays a vital role for having a reasonable amount of lepton asymmetry which accounts for successful baryogenesis. For three flavored leptogenesis we present the variation BAU with respect to the model parameters in Fig \ref{fig8}. 

\clearpage
\section{Conclusion}
\label{sec5}
We have exercised a detailed analysis on baryogenesis via leptogenesis scenario considering a framework augmented with $A_4$ flavor symmetry where type I seesaw has been chosen as the mechanism of neutrino mass generation. Here we have considered two vev alignments for the extra flavon $\eta$ one kind of which allow us to accommodate $\eta$ as a stable dark matter candidate by respecting the $Z_2$ symmetry. As already mentioned, here we have only shown the results of BAU with respect to the light neutrino model parameters since neutrino phenomenology has already been explored in the work \cite{Meloni:2010sk}. While searching for the allowed model parameter space, we have used the latest global fit neutrino oscillation data for two mass squared splittings in their allowed $3\sigma$ range and found relations among the various model parameters and the known light neutrino parameters. With the help of those light neutrino model parameters we find baryon asymmetry. In order to find the same we have chosen three ranges for the RHN mass as necessary for the corresponding flavor regimes. We have checked the viability of both flavored and unflavored leptogenesis by considering different possible mass regimes concerned with the right handed neutrinos. It is clear from the results that the model addressed leptogenesis at all the possible scales of RHN mass as mentioned in subsection A of Sec.\ref{sec3}. Of special importance is the fact that the RHN mass squared ratio ($\text{x}_j$) for a particular mass regime plays a crucial role in achieving an adequate amount of lepton asymmetry which accounts for the observed matter-antimatter asymmetry. We presented a table \ref{tab3} summarizing the various RHN mass scales and the mass squared ratios which are demanded by the model under consideration in order to explain matter-antimatter asymmetry. Some important observation can be made from the results and analysis:
\begin{itemize}
\item For all the flavor regimes e.g. one, two and three the model prediction meets the observed BAU.
\item To account for a required amount of lepton asymmetry for one flavor regime the mass splitting $\text{x}_j$ can be a bit larger as compared to that required for two flavor and/or three flavor.
\item With the same mass splitting $\text{x}_{j}$ as kept for one flavor regime, it is difficult to reproduce an adequate amount of lepton asymmetry for two flavor regime. As we go below $10^{12}$GeV for RHN mass which is a criteria for having two/three flavor leptogenesis the factor $\text{x}_j$ needs to be a little less to have an enhanced lepton asymmetry in order to generate the observed BAU.
\item Then for fully flavored regime or three flavor leptogenesis we choose the RHN mass scale to be around $10^8$ GeV. With this RHN mass, the mass squared ratio($\text{x}_j$) requires much more smaller value to have a sufficient amount of lepton asymmetry to meet the observed BAU as reported by Plank 2015 data.
\item The common parameter space for nonzero reactor mixing angle as envisaged by the model and baryon asymmetry is difficult to coincide in case of one flavor leptogenesis, whereas in case of two and three flavored leptogenesis, the plot   between baryon asymmetry versus $\text{Sin} ^2 \theta_{13}$ says that their parameter space matches only for a very narrow region. This fact allows us to have more predictability of the RHN mass scale relevant to that particular range required for two and/or three flavored leptogenesis.
\end{itemize}

Notwithstanding the prediction for experimentally observed neutrino parameters, this model beautifully sheds light on one of the long standing puzzle of particle physics and cosmology, the baryon asymmetry of the universe. The model also touches the dark sector accommodating a stable dark matter candidate, the detailed exercise of which we keep for our next draft. 
 
\appendix
\section{$A_4$ product rules}
\label{appen1}
$A_4$ is isomorphic to the symmetry group of a tetrahedron. It is a discrete non-Abelian group of even permutations of four objects. It has four irreducible representations: three one-dimensional and one three dimensional which are denoted by $\bf{1}, \bf{1'}, \bf{1''}$ and $\bf{3}$ respectively. Their product rules are given as
$$ \bf{1} \otimes \bf{1} = \bf{1}$$
$$ \bf{1'}\otimes \bf{1'} = \bf{1''}$$
$$ \bf{1'} \otimes \bf{1''} = \bf{1} $$
$$ \bf{1''} \otimes \bf{1''} = \bf{1'}$$
$$ \bf{3} \otimes \bf{3} = \bf{1} \otimes \bf{1'} \otimes \bf{1''} \otimes \bf{3}_a \otimes \bf{3}_s $$. Presenting two triplets as $(a_1, b_1, c_1)$ and $(a_2, b_2, c_2)$ respectively, their direct product can be written as
$$ \bf{1} \backsim a_1a_2+b_1c_2+c_1b_2$$
$$ \bf{1'} \backsim c_1c_2+a_1b_2+b_1a_2$$
$$ \bf{1''} \backsim b_1b_2+c_1a_2+a_1c_2$$
$$\bf{3}_s \backsim (2a_1a_2-b_1c_2-c_1b_2, 2c_1c_2-a_1b_2-b_1a_2, 2b_1b_2-a_1c_2-c_1a_2)$$
$$ \bf{3}_a \backsim (b_1c_2-c_1b_2, a_1b_2-b_1a_2, c_1a_2-a_1c_2)$$

\bibliographystyle{apsrev}

\end{document}